# Community Detection in Complex Networks Using Agents

Ismail Gunes
Bogazici University
Dept. of Computer Engineering
Bebek, Istanbul, Turkey

ismail@uekae.tubitak.gov.tr

Haluk Bingol
Bogazici University
Dept. of Computer Engineering
Bebek, Istanbul, Turkey

bingol@boun.edu.tr

## ABSTRACT
Community structure identification has been one of the most popular research areas in recent years due to its applicability to the wide scale of disciplines. To detect communities in varied topics, there have been many algorithms proposed so far. However, most of them still have some drawbacks to be addressed. In this paper, we present an agent-based based community detection algorithm. The algorithm that is a stochastic one makes use of agents by forcing them to perform biased moves in a smart way. Using the information collected by the traverses of these agents in the network, the network structure is revealed. Also, the network modularity is used for determining the number of communities. Our algorithm removes the need for prior knowledge about the network such as number of the communities or any threshold values. Furthermore, the definite community structure is provided as a result instead of giving some structures requiring further processes. Besides, the computational and time costs are optimized because of using thread like working agents. The algorithm is tested on three network data of different types and sizes named Zachary karate club, college football and political books. For all three networks, the real network structures are identified in almost every run.

## Categories and Subject Descriptors
I.2.11 [**Artificial Intelligence**]: Distributed Artificial Intelligence – *Multiagent systems*

## General Terms
Algorithms

## 1. INTRODUCTION
There has recently been an increasing interest on the use of complex networks in various research areas. The complex network is the representation of a system in which there are interactions between its components constantly evolve and unfold over time [1]. A complex network is represented by the graph theory; it has nodes as members and edges linking them according to some relationship. The most known examples of the complex networks are the World Wide Web (WWW), internet, actor collaboration networks and metabolic networks. For instance, the nodes of the network are web pages in the network of WWW and the edges are the hyperlinks from one page to another.

The most of the complex networks have vertices in a group structure that the vertices within the groups have higher density of edges while vertices between groups have lower density of edges [5]. This structure is called *community structure*. The communities in a network may reveal many hidden features of the network such as relationships between the members or the organizations among members of subgroups in the network. The nodes belonging to the same community are more probable to have properties in common [3]. For example, the communities denote the thematic clusters in the WWW. In metabolic networks, the communities may represent the functional groups and detecting these groups may simplify the functional analysis in biochemistry. Also, the information flow mechanism of the network may be revealed by the community structure. Today, detecting communities is considered to be used in advertising, detecting terrorist organizations in the WWW etc. Hence, detecting community structure may be an indispensable part of the complex system analysis, in upcoming years.

In order to learn about the community structure of the network, the software agents are used in our algorithm. A software agent is a software entity that acts with autonomy to accomplish tasks on behalf of its users. They function continuously and autonomously in a particular environment, often inhabited by other agents or processes [8]. Although the autonomous and collaborative function of the agents is appropriate for the community detection, it is rarely used to identify the communities. One of these few algorithms is the Young et al. [10]. They have proposed an agent-based algorithm for detecting communities in the networks and this is the first paper using agents in community detection. However, the algorithm has some drawbacks. Our algorithm makes many developments on this algorithm and makes the algorithm self-contained.

Before detecting the community structure in a network, it is oblige to define what a community is. There are many researches trying to find a measurement for assigning a subgroup as the community and find the level of the community structure. The most widely used community definition metric in partitioning a network into communities was proposed by the Girvan and the Newman [6]. The metric is called *the network modularity* and represented as;

$$Q = \sum_i \left(e_{ii} - a_i^2\right)$$

where, $i$ is the index of the communities, $e_{ii}$ is the fraction of edges, that connects two nodes inside the community $i$, to the total number of edges in the network and $a_i$ is the fraction of all the edges whose at least one node in the community $i$ to the total number of edges in the network. If the edges are similar with the random connections, the modularity value will be close to the 0. However, the modularity value close to the 1 denotes the strong community structure. In practice, the modularity value is expected to be in 0.3 and 0.7. The community structure providing the maximum network modularity value is selected in our algorithm.

In this paper, we propose an agent-based algorithm for detecting the community structure in the networks, which makes agents perform biased moves in a smart way. The algorithm also uses the network modularity to determine the number of communities. The need for the prior knowledge about the network such as number of the communities or any threshold value is removed in the algorithm. Besides, the absolute community structure is given as a result instead of a *dendrogram* requiring further process.

## 2. COMMUNITY DETECTION METHODS

The community detection has attracted great interest recently because of many benefits behind it. The nodes included in the same community are assumed to have similar properties. By identifying the communities, not only these similarities are found but also the network structure may be revealed. The most well-known instance is about World Wide Web (WWW). Among millions of web pages, it is hard to determine which websites are common in subject and contains similar materials. The information about these clusters is important and may be used for different areas such as advertising, marketing etc. Furthermore, the communities may be used in the metabolic networks in order to divide the network into functional subgroups. Therefore, detecting communities may be used for uncovering the operations and transactions in the network.

There have been many different approaches and algorithms to detect the community structure in complex networks. The most known algorithm is the Girvan-Newman (GN) algorithm [10] that is based on the betweenness centrality. The algorithm is a divisive method and includes the removal of the edges depending on their betweenness instead of removing the edges joining the nodes with lower similarities. The betweenness value may be defined in different ways depends on the application. At last, the algorithm produces a hierarchical structure of the network that is called *dendrogram.* The most distinguishing part of the GN algorithm is calculating the betweenness values after each removal and providing consistent results. The GN algorithm works well with the networks with known community structure. However, it requires high computational effort which is not feasible for large networks.

The algorithm proposed by the Clauset et al. [2] is feasible for detecting communities in very large networks. It is a hierarchical agglomerative method that uses greedy optimization. To find the division of the network with highest modularity, the candidates are searched by tracking the changes in the modularity values. Because of this optimization phase, the algorithm may be applicable to very large networks. This algorithm also produces a dendrogram and defines a way to cut the dendrogram at some level.

The Radicchi's algorithm [7] is similar with the GN algorithm but it resolves some issues not addressed in prior algorithms. First, it proposes a new formula to define the communities and removes the need for a prior knowledge about the network. The algorithm considers triangles in the network while detecting the communities and makes algorithm self-contained. Another issue is reducing the computational effort of the GN algorithm. To perform this, only local quantities are used instead of calculating all edge betweenness values. Hence, the Radicchi's algorithm provides accurate results as GN algorithm with lower computational cost.

Another algorithm benefits from the network modularity is the external optimization algorithm of Duch and Arenas [4]. It is also a divisive algorithm that uses a heuristic search on the external optimization. It is based on optimizing a global variable by developing contributions of local variables. The algorithm tries to optimize the modularity and defines a fitness value. The network is divided into two partitions with equal elements and a node with the lowest fitness value is moved to the other side at each step. This operation is continued until there is no improvement in the modularity value. This algorithm provides higher modularity values than the others but it has higher time complexity. So, it is not efficient for large networks.

### 2.1 Drawbacks of current algorithms

Although there are many algorithms which have been proposed so far, most of them have some common drawbacks. First, some community detection algorithms need priori knowledge about the network. This priori knowledge may be the desired number of the communities or any threshold values. There may also be some restrictions about community sizes. However, the algorithms with these limitations may not be applicable to the networks in the real life.

Another common drawback in the community detection algorithms is the high computational or time cost. The algorithms providing accurate results have higher costs in general. However, not only the accurate results are needed in community detection but also acceptable time and computational complexities. Today, the fastest algorithm runs in linear time [9] but this algorithm has an assumption that all communities are of the similar sizes and it needs some knowledge about number of the community number.

## 3. THE ALGORITHM

Our algorithm is based on finding community structure using smart agents in the complex networks. As the algorithm works, the agents learn more about the network by the time and perform better. Hence, we call them *smart agents.*

Although the software agents are rarely used in community detection, they are appropriate for network analysis issues. The main reason behind the idea of using agents in the community detection is their collaboration. The software agents function together and work in a coordinative way. That is, lots of agents are used in the network and the global structure of a network may be revealed. Besides, the easy messaging mechanism of the agents may be used while detecting communities in a network.

The algorithm we propose is a stochastic one. That is, the same input may not give the same results for different runs of algorithm. The operation of detecting communities has not an absolute solution or strict boundaries. Hence, it may be addressed by a stochastic approach. Although this indeterminist behavior usually provides accurate results, this stochastic approach is open to discussions. The details of the algorithm will be given in the next subsections

## 3.1 Algorithm Overview

There are two main phases in the algorithm. The first phase is the collective exploration and it is handled by the smart agents. In the former phase, the agents traverse in the graph in a biased way and collect information about current network structure. This collected information is analyzed and the community structure is found using the modularity values in the latter phase.

The proposed algorithm that runs on undirected graph needs no priori knowledge about the network. There are many algorithms like GN algorithm that provides a dendrogram at the end of the algorithm and needs a value to decide on the communities. Unlike these algorithms, our algorithm provides definite community structure and there is no need for any further processes.

The main idea behind using agents in community detection is the tendency of agents to stay in the community rather than leaving the community. A community in the network is defined to be a densely connected group of vertices. Hence, there are more edges with both starting and ending vertices within community than the edges with vertices outside the community. So, it is more probable for an agent to stay in the community because there are more ways for the agent to stay within the community than there are to leave it. To clarify this idea, gas molecules metaphor can be used. Consider two rooms connected by a door. A molecule in one of the rooms is randomly moving around. As long as the door is small, the molecule is tent to stay in the room it started with. Likewise, the communities in a network may be thought as the rooms connected with some doors. So it is more probable for an agent to stay within the community it started with.

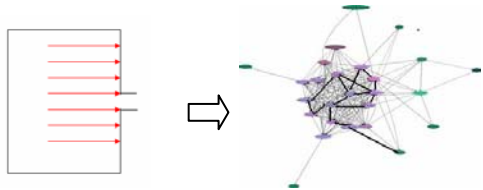

**Figure 1. The idea behind using agents in community detection**

## 3.2 The Exploration Phase

The exploration phase is the first part of the program and the network structure is explored in this part. The idea about how may be the community structure of the network is captured in this part. Hence, the exploration phase should be handled accurately to make analysis phase work well. The information about network structure which will be used in the analysis phase is kept in a matrix structure after this phase.

There are two types of agents in the algorithm. The information about the network structure is collected by lots of agents of type *slave agents* and evaluated by only one agent of type *coordinator agent*. The other agents travel in the network and send visited nodes to the main agent. The coordinator agent receives these feedbacks and updates the *weight matrix* that will keep the biases of a traverse operation from a node to the neighbors. After that, the coordinator agent will have a global knowledge of the network coming from different parts of the network. The learning of slave agents is realized by these update operations.

At first, the coordinator agent gets the network data into the adjacency matrix that holds the neighborhood information and initiates the slave agents. The slave agents work iteratively in a cyclic behavior and the coordinator agent also listens to the feedback messages from the slave agents, continuously with cyclic behavior.

The first generation agents of type slave agents behave completely random. They are placed on different vertices which are selected randomly. Then, they start to travel in a biased way that the most recently visited nodes are not visited as much as it is possible. Each agent has its own *memory*. During traveling in the graph, the traversed nodes are stored in the memory. After traversing certain number of nodes, the nodes stored in this memory are sent to the coordinator agent in a *propose message* to be evaluated.

The coordinator agent and the slave agents work asynchronously using the messages. Not only the hand-shaking is provided by the messages, but also informing the coordinator agent about traveled nodes. If the incoming message from slave agents to the coordinator agent is a *propose message* that reports the traveled nodes by an agent in a number of hop counts (*memory* of an agent), the traveled nodes are obtained by the coordinator agent. After getting the traveled nodes, the weight matrix that initially consists of zeros is updated. The weight matrix is an n ✕ n matrix where n is the number of nodes. The message containing the traveled nodes and the related update operation in the coordinator

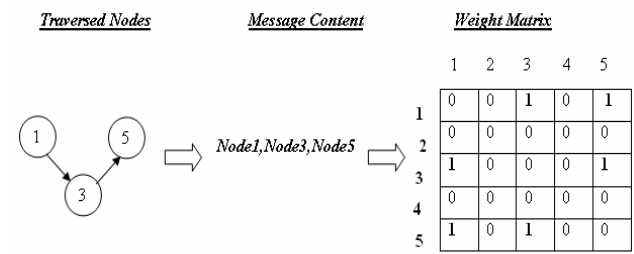

agent is figured below.

**Figure 2. The travel of an agent, reporting traveled nodes and related update operation in the main agent.**

The nodes included in the same message (memory) from an agent are assumed to be *probably near*. Then, the weight values between these probably near nodes are incremented.

After updating the weight matrix, the next generation agents of type slave agents start traversing again and make their moves according to the new weight matrix. Hence, slave agents behave reasonable instead of behaving randomly. The update operations provide some improvements; increase accuracy and slave agents consider the new weight matrix afterwards. At last, we can say that these update operations by the coordinator agent make the slave agents learn and change them from agents into *smart agents*.

The next generation agents are located in the vertices with most hits in the graph, instead of randomly selected vertices. The reason behind assigning the hub nodes as starting points is strengthening the idea of staying in the community and better exploration of communities. However, assigning only hub nodes as starting points may prevent discovering all parts of the graph. Therefore, some least visited nodes are also assigned as starting points. Hence, different parts of the network are also visited and they participate into the competition to be a hub node.

After selecting the starting nodes, the next generation agents are ready to travel in the graph. While traveling, they consider the weight matrix and able to make reasonable moves instead of random ones. The *probably near* neighbors of a node are found and the more chance is given to these nodes for traveling. Using *weight matrix,* the weight values for each edge with a node and its neighbors are found. To give the edges with 0 weights a chance to be used, the weight in the formula is incremented by 1. The probability of using an edge for a biased move is calculated like below.

$$P(j) = \sum_j \frac{(1+w_j)}{(1+w_i)}$$

where, $w_j$ denotes the weight value of edge between a node and its $j^{th}$ neighbor and $p(j)$ denotes the probability of moving to the $j^{th}$ neighbor. After these traverse operations, the agents send the traversed nodes to the coordinator agent again and these data is also used for updating the weight matrix. The later generations work in the same way and network structure exploration is optimized as algorithm goes on.

The exploration phase is terminated when it is decided that the whole graph is explored enough. To decide, each node should be visited at least (number of agents - 1) ✘ (size of agent memory) times.

## 3.3 The Analysis Phase
After the exploration phase is handled by the slave agents, the collected information in the weight matrix from the exploration phase is used in the analysis phase by the coordinator agent. Then, the edges with the lowest weight values are removed from the graph. After removing the edge with the lowest weight from the graph, it is checked whether a new community is formed or not. To check this, an algorithm called "Flood Fill", available at http://www.comp.nus.edu.sg/~stevenha/, that gives the number of connected components in a graph is used. The basic idea here is finding the node which has not been assigned to a component yet and finding the component that contains it.

Using this algorithm, we find out which edges to be removed to produce which community structure. If removing an edge produces a new connected component, these components are considered to be the candidate communities. Then, the modularity value is calculated to select among these candidates to find the real community structure. The candidate connected components giving the highest modularity value is assumed to be the community structure of the network.

## 4. EXPERIMENTAL RESULTS
Our algorithm was tested in three different datasets. The only parameters varying according to the dataset are the number of agents and the size of agent memories to be used. A few numbers of agents are used for small datasets, while more agents are used for bigger datasets. The datasets of different sizes we have used are the Zachary Karate Club, College football league and the Political Books datasets.

## 4.1 Zachary Karate Club
The Zachary Karate Club data [11] is collected by Wayne Zachary from the karate club of a university. This dataset is widely used in testing of community detection algorithms. The vertices are the students and there is an edge between two vertices if these two students are good friends. The two communities represent two karate clubs which are formed after a disagreement. This network includes 34 nodes and 78 edges. Our algorithm is run on this dataset many times and at least 33 nodes from 34 nodes are placed in right communities while a node is misplaced sometimes. Our algorithm provides 97% accurate results at least with this dataset.

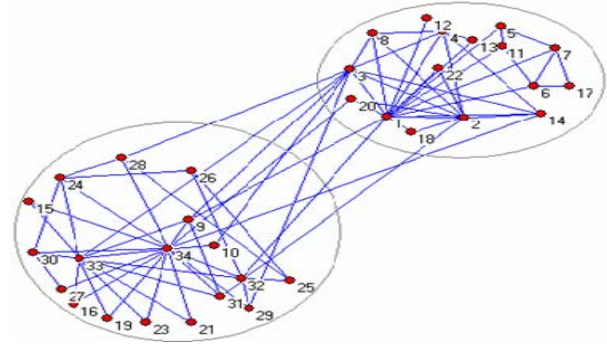

**Figure 3. Real community structure of Zachary Karate Club**

## 4.2 College Football Network
The College football data is collected using the football matches in USA during year 2000. The nodes represent the teams in the conferences and there is an edge between two teams if they played a match at least once. Teams are allowed matches with teams from other conferences but the probability of making matches within conference is higher. The community detection we want to find is the conference structure. The college football network dataset contains 93 teams as nodes and 452 matches as edges. There are 10 conferences and these conferences show the real community structure of the network. That is, there are 10 communities to detect. Our algorithm finds out all 10 communities for most of runs and produces 95% accurate results at least.

## 4.3 Political Books Network
The political books data is produced by Valdis Krebs, books dataset available at http://www.orgnet.com/leftright.html, in 2003 using the book buying information on some retailers. The information like "Customers who bought this book also bought:" on the web pages is used to form the network. The nodes denote the political books bought and there is an edge between two nodes if they were bought together at a major retailer on the web. The books are divided into two different categories depending on the idea they contain. These two categories should be found as the community structure. This network of political books has 49 nodes and 292 edges. Our algorithm is run on this dataset several times and three communities are detected in every time. The two communities including different ideas are definitely detected and the book that does not belonging to both ideas is also detected as a separate community. Hence, the algorithm provides 100% accurate results with this network.

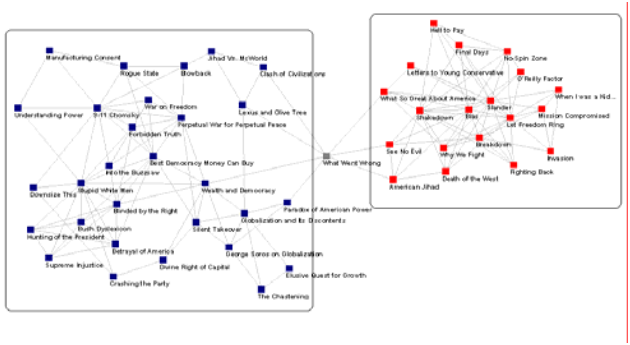

**Figure 4. Real community structure of Political Books network**

## 5. CONCLUSIONS

In this paper, we have proposed a new community detection algorithm, which uses agents to explore the community structure. Then, the modularity value is calculated for determining on the number of communities. Unlike most of the community detection algorithms, our algorithm needs no priori knowledge about the network like number of communities, threshold value or size of the communities. Moreover, our algorithm produces the definite communities and the members instead of a dendrogram. Hence, there is no need for any further process. Last, the algorithm provides accurate results with acceptable time cost due to the asynchronously working agents.

Our algorithm has some contributions to the rarely use of agents in community detection; i) the next generations are used addition to the first generation agents, ii) the agents are placed in the network depending on the hit numbers of the nodes instead of placing randomly, iii) the agents move in a smart way that they use a global knowledge about the network for a biased move. Hence, they produce better results by the time, iv) the algorithm needs no prior knowledge like the number of communities, v) there is no need for the voting cut off parameter or a clean up phase.

We have tested our algorithm on different datasets of different sizes and subjects. For all three datasets, the algorithm produces reasonable results. At least 95% accurate results are provided for these datasets and the real community structure is explored in reasonable time.

As a future work, the algorithm may be run in a distributed environment. The agents work on different computers and the main agent works on a computer with higher capacity to perform central evaluations. By distributing the work load, the algorithm may be more efficient. The coordination and the easy messaging mechanism make agents appropriate for distributed working. Another addition to the algorithm would be the learning of the number of agents and the size of the agent memories depending on the dataset by the time. Last, more and larger datasets may be used for testing the algorithm.

## 6. ACKNOWLEDGMENTS

This work was partially supported by Bogazici University Research Projects under the grant number 06A105.